\documentclass[useAMS,usenatbib,letterpaper]{mnras}
\usepackage{epsfig,amsmath,amssymb,natbib}
\usepackage{aas_macros}

\begin{document}

\title[Constraining $\sigma_8$ and $b$ in Sloan]{\LARGE Galaxy Bias and $\sigma_8$ from Counts in Cells from the SDSS Main Sample}

\author[A. Repp \& I. Szapudi]{Andrew Repp\ \& Istv\'an Szapudi\\Institute for Astronomy, University of Hawaii, 2680 Woodlawn Drive, Honolulu, HI 96822, USA}

\date{\today; to be submitted to MNRAS}

\label{firstpage}
\pagerange{\pageref{firstpage}--\pageref{lastpage}}
\maketitle

\begin{abstract}
The counts-in-cells (CIC) galaxy probability distribution depends on both the dark matter clustering amplitude $\sigma_8$ and the  galaxy bias $b$. We present a theory for the CIC distribution based on a previous prescription of the underlying dark matter distribution and a linear volume transformation to redshift space. We show that, unlike the power spectrum, the CIC distribution breaks the degeneracy between $\sigma_8$ and $b$ on scales large enough that both bias and redshift distortions are still linear; thus we obtain a simultaneous fit for both parameters. We first validate the technique on the Millennium Simulation and then apply it to the SDSS Main Galaxy Sample. We find  $\sigma_8 = 0.94^{+.11}_{-.10}$ and $b = 1.36^{+.14}_{-.11}$, consistent with previous complementary results from redshift distortions and from Planck.

\end{abstract}


\section{Introduction}
\label{sec:intro}
Galaxy surveys are a principal means of constraining cosmology; however, analysis of these surveys must account for the fact that galaxies are biased tracers of matter. On large (``linear'') scales -- typically $30h^{-1}$Mpc or larger -- one can parametrize this bias using a simple first-order expansion of the relationship between galaxy- and matter-densities. (see theoretical foundation in \citealp{Kaiser1984}, elaborated in \citealp{BBKS}). In particular, if $\delta_g$ and $\delta$ are the galaxy- and matter-overdensities, respectively, linear bias theory assumes $\delta_g = b \delta$; as a result, $b^2$ is the ratio (constant in this theory) between the galaxy and dark matter power spectra.

Since this bias shifts the galaxy spectrum by a simple multiplicative factor, it is in linear theory completely degenerate with the amplitude parameter $\sigma_8$. Methods of breaking this degeneracy include cross-correlation with CMB lensing \citep{Baxter2019}, analysis of redshift space distortions (e.g., \citealp{Howlett2015}), and consideration of higher-order statistics (e.g., \citealp{Szapudi1998, PanSzapudi2005}).

We here demonstrate a complementary method of breaking this degeneracy, namely, fitting the measured probability distribution of galaxy counts. Section~\ref{sec:pdffit} outlines both the theory behind this method and our implementation of it, showing that the method successfully recovers $\sigma_8$ and $b$ in the Millennium Simulation. Section~\ref{sec:data} describes the SDSS galaxy data which we proceed to fit; our results appear in Section~\ref{sec:results}, and our conclusions in Section~\ref{sec:concl}.

\section{Fitting the One-Point Probability Distribution}
\label{sec:pdffit}
\citet{Uhlemann2020} show that simultaneous fits to both the matter power spectrum $P(k)$ and the one-point probability distribution function (PDF) yield significantly tighter cosmological constraints than analysis of the power spectrum alone. Thus the matter PDF contains information which the power spectrum does not.

Turning to the galaxy PDF, its second moment $\sigma_g^2$ depends on the amplitude of the galaxy power spectrum. Since (in the linear regime) this spectrum is proportional to the square of $b\sigma_8$, it follows that $\sigma_8$ and the linear galaxy bias $b$ are completely degenerate in their effect on the second moment of the PDF. The third moment ($S_{3g}$), however, depends on $1/b$ but not on $\sigma_8$ \citep{FryGaztanaga1993,Juszkiewicz1995}; thus, the full probability distribution contains enough information to break the degeneracy between the two parameters, as noted by \citet{SzapudiPan2004}.

Hence, the following sections outline our application of PDF-fitting to galaxy surveys. Section~\ref{sec:pdffittheo} presents the theoretical foundation of the method; Section~\ref{sec:pdffitdesc} explains the choices and assumptions behind our implementation of it; and Section~\ref{sec:pdffitval} reports its validation against the Millennium Simulation.

\subsection{Theory}
\label{sec:pdffittheo}
Local models of galaxy bias express the number of galaxies $N$ as a function of the underlying dark matter density in a survey cell. For a linear bias model, the galaxy overdensity $\delta_g = N/\overline{N} - 1$ is a constant multiple of the dark matter overdensity $\delta = \rho/\overline{\rho} - 1$ ($\rho$ being the dark matter density in the survey cell): i.e., $\delta_g = b \delta$, so that $N = \overline{N} \left(b\delta + 1 \right)$.

One can easily incorporate stochasticity by taking the output of the bias model as the expected value of $N$ given $\delta$; for linear bias,
\begin{equation}
\langle N \rangle_\delta = \overline{N} \left(b\delta + 1 \right).
\label{eq:linbias}
\end{equation}
In our case, however, one requires additional specification of the conditional distribution $\mathcal{P}(N|\delta)$ in order to obtain  the one-point probability distribution of galaxy counts-in-cells:
\begin{equation}
\mathcal{P}(N) = \int d\delta \mathcal{P}(\delta) \mathcal{P}\left(N|\delta\right),
\end{equation}
where $\mathcal{P}(\delta)$ is the underlying dark matter distribution.

Finally, galaxy surveys typically employ redshift as a proxy for radial distance; as a result, any radial motion of galaxies with respect to the Hubble flow distorts the measured distance to those galaxies. On the linear scales relevant to this work, the predominant source of radial motion is the coherent infall of galaxies into overdense regions; the resulting distortion is equivalent (e.g., \citealp{Szapudi2004}) to a volume transformation. Using \citet{Kaiser1987}'s analysis one can show that the resulting (monopole) redshift-space galaxy power spectrum is
\begin{align}
P^{(s)}_g(k) & = \left( 1 + \frac{2}{3}\beta + \frac{1}{5}\beta^2 \right) P_g(k),\\
                   & =  \left( 1 + \frac{2}{3}\beta + \frac{1}{5}\beta^2 \right) b^2 P(k)
\end{align}
where $\beta = f/b$ depends upon both the linear growth rate $f \approx \Omega_m^{0.55}$ and the linear galaxy bias $b$.

By integrating this spectrum with the proper smoothing kernel, we can obtain the redshift-space variance of the galaxy number counts for a given smoothing radius $R$:
\begin{align}
\sigma^2_{(s),g\,R} & = \left( 1 + \frac{2}{3}\beta + \frac{1}{5}\beta^2 \right) b^2 \, \sigma^2_R\\
 & = \left( 1 + \frac{2}{3}\beta + \frac{1}{5}\beta^2 \right) b^2 \, (K\sigma^2_8),\label{eq:zspsig}
\end{align}
where $K$ is a proportionality constant. In real space, on the other hand, we have
\begin{equation}
\sigma^2_{g\,R} = b^2 \, \sigma^2_R = b^2 \, (K\sigma^2_8).\label{eq:Rspsig}
\end{equation}
Comparing Equations~\ref{eq:zspsig} and \ref{eq:Rspsig}, we can define an effective redshift-space value of $\sigma_8$ as follows:
\begin{equation}
\sigma_{8,\mathrm{eff}} = \sigma_8 \sqrt{1 + \frac{2}{3}\beta + \frac{1}{5}\beta^2 }.
\label{eq:sig8eff}
\end{equation}
Since higher-order moments of the PDF are relatively insensitive to redshift-space distortions \citep{Hivon1995}, the change from $\sigma_8$ to $\sigma_{8,\mathrm{eff}}$ accounts for the bulk of the effect on the galaxy probability distribution. Thus, as long as we remain in the linear regime, we can fit $\sigma_{8,\mathrm{eff}}$ and $b$ to the redshift-space galaxy PDF and then use Equation~\ref{eq:sig8eff} to obtain the true underlying $\sigma_8$. 

In summary, to fit cosmological and bias parameters to an observed set of galaxy counts, one requires a prescription for the dark matter distribution (depending on cosmology in general and $\sigma_8$ in particular); a galaxy bias model to provide $\langle N \rangle_\delta$; and further specification of a conditional probability model $\mathcal{P}(N|\delta)$ for stochastic galaxy formation. In the subsequent section we turn to the specification of these three ingredients.

\subsection{Implementation}
\label{sec:pdffitdesc}
Central to the implementation of this method is a prescription for the one-point dark matter distribution. Note that mere fits to the output of a specific simulation do not suffice for this purpose unless the fits are generalized in terms of cosmological parameters.

One such prescription, based upon large deviation statistics and spherical collapse, is that of \citet{Uhlemann2020}. This theoretically-motivated prescription is generally applicable to smoothing radii as small as $10h^{-1}$Mpc.

Another (more phenomenological) prescription appears in \citet{ReppApdf}, employing the Generalized Extreme Value (GEV) distribution and based on rescalings of the Millennium Simulation \citep{Springel2005, Angulo2012}:
\begin{equation}
\label{eq:GEV}
\mathcal{P}(\delta) = \frac{1}{(1+\delta)\sigma_G} t(\delta)^{1+\xi} e^{-t(\delta)},
\end{equation}
where
\begin{equation}
\label{eq:GEV_t}
t(\delta) = \left(1 + \frac{\ln \delta - \mu_G}{\sigma_G}\xi\right)^{-1/\xi}.
\end{equation}
Here, $\mu_G$, $\sigma_G$, and $\xi$ are location, scale, and shape parameters (respectively), with cosmology-dependence as outlined in \citet{ReppApdf}. This GEV prescription applies to dark matter densities in cubical cells as small as $2h^{-1}$Mpc, and it produces a cumulative distribution function with better than 2 per cent accuracy. \citet{Klypin2018} have shown that it overpredicts the (already quite low) probability at very large densities ($\delta \ga 90$); however, for the cell sizes in this work, the data sets we analyze remain well below this threshold ($\delta_g^\mathrm{max} < 5$). Therefore, we use the GEV dark matter PDF in this work and ignore this high-end inaccuracy.

For our galaxy bias model we use the linear bias of Equation~\ref{eq:linbias}. Although the model becomes inaccurate and even unphysical on small scales \citep{Ising1}, on linear scales ($\ell \ga 30h^{-1}$Mpc) it seems to provide a good fit to simulations and observation and furthermore receives support from perturbation theory \citep{Desjacques2018}. Thus, to remain in the linear model's regime of applicability, we employ cubical cells of $\sim 32h^{-1}$-Mpc sides.

Finally, although the bias model yields a mean number of galaxies for a given $\delta$, it is evident that many factors besides the dark matter density influence the galaxy formation process. Thus we require additional assumptions -- such as Poisson statistics -- for converting the mean $\langle N \rangle_\delta$ into a probability distribution $\mathcal{P}(N|\delta)$. We note that \citet{Gruen2018} have shown that Dark Energy Survey (DES) data indicate stochasticity (beyond Poisson scatter) in $20'$ apertures within a photometric redshift range of 0.2--0.45; they note however that on larger scales ($\ga 10h^{-1}$Mpc) such additional stochasticity is expected to be small. 
Since we are working at these larger scales, this work assumes Poisson scatter: $\mathcal{P}(N|\delta) = \mathrm{Pois}\left(N | \langle N \rangle_\delta \right)$.

Summarizing, we use Equations~\ref{eq:GEV} and \ref{eq:GEV_t} for $\mathcal{P}(\delta)$, handling redshift-space distortion by Equation~\ref{eq:sig8eff}; and we assume a linear bias model to obtain $\langle N \rangle_\delta$, with Poisson scatter yielding $\mathcal{P}(N|\delta)$.

\subsection{Validation}
\label{sec:pdffitval}
To validate the PDF-fitting method of the previous two section, we  first apply it to simulated results. For this purpose we use the galaxy catalog described in \citet{Bertone2007}, obtained by application of the L-Galaxies semianalytic model (Munich model, \citealp{Croton2006, deLucia2006}) to the Millennium Simulation \citep{Springel2005}.\footnote{Galaxy catalog downloaded from the repository at \texttt{http://gavo.mpa-garching.mpg.de/Millennium/}} To obtain a uniform galaxy sample, we require stellar mass $M_\star \ge 10^9 \mathrm{M}_\odot.$ Using the plane-parallel approximation, we then prepare a redshift-space catalog by shifting the $z$-coordinate of each galaxy by a factor corresponding to $v_z$, the third component of the galaxy's proper motion.

At this point we divide the simulation volume of ($500h^{-1}$Mpc)$^3$ into $16^3$ cubical cells, each with side length $31.25h^{-1}$Mpc, counting the galaxies in each cell. To obtain the distribution $\mathcal{P}(N)$, we choose 20 bins (logarithmic in $N$) such that the first bin contains only the lowest $N$-value (in this case, $N_\mathrm{low} = 191$).  We then combine bins (starting with the lowest) as necessary so that no bin contains fewer than three survey cells, ending up with 18 final probability bins.

We estimate the probability $\hat{\mathcal{P}}(N)$ of measuring $N$ galaxies in a cell, where $N$ falls within a bin $B$ of (integer) width $\Delta N$. Let us denote with $n_c$ the number of cells containing $N$ galaxies, out of $N_\mathrm{tot}$ total cells; then 
\begin{equation}
\sum_{N \in B} n_c(N)/N_\mathrm{tot} = \mathcal{P}\left(\lbrace N | N \in B \rbrace\right) \equiv P_B,
\end{equation}
so that for any single $N$ in the bin we have
\begin{equation}
\hat{\mathcal{P}}(N) = \frac{\sum_{N \in \mathrm{bin}} n_c(N)}{N_\mathrm{tot} \cdot \Delta N} = \frac{P_B}{\Delta N}.
\label{eq:PNest}
\end{equation}
We center our counts-in-cells estimator in each bin on $\hat{N} = \langle N \rangle_B$.
We also need $\hat{\sigma}^2_{\mathcal{P}(N)}$, which we estimate as follows: fixing a value of $N$, we assign to each survey cell a value of $\mathcal{S} = 1$ if it contains $N$ galaxies and $\mathcal{S} = 0$ otherwise. Then the first two moments of $\mathcal{S}$ are $\langle \mathcal{S} \rangle = \langle \mathcal{S}^2 \rangle = \mathcal{P}(N)$, so that $\sigma_\mathcal{S}^2 = \mathcal{P}(N) - \mathcal{P}(N)^2.$ But if $N_\mathrm{tot}$ is the total number of cells in the survey, then we have $N_\mathrm{tot}$ measurements of $\mathcal{S}$. Furthermore, at $32h^{-1}$-Mpc scales the correlation between survey cells -- and thus between these measurements of $\mathcal{S}$ -- is negligible; hence for the error on $\mathcal{P}(N)$ we write 
\begin{equation}
\sigma_{\mathcal{P}(N)}^2 = \sigma_{\langle \mathcal{S} \rangle}^2 = \frac{\sigma_\mathcal{S}^2}{N_\mathrm{tot}} = \frac{\mathcal{P}(N) \left( 1 - \mathcal{P}(N) \right)}{N_\mathrm{tot}}.\label{eq:sigPN}
\end{equation}
When using a binned version of the CIC distribution, we essentially have $\Delta N$ independent measurements of $\hat{\mathcal{P}}(N)$ (cf. Equation~\ref{eq:PNest}); thus we must divide the values from Equation~\ref{eq:sigPN} by $\Delta N$ to obtain final error $\hat{\sigma}^2_{\mathcal{P}(N)}$. This is the final error that we will use in our fit.

Before proceding to the fit, we note that the scale at which we are performing the measurements ($\sim 30h^{-1}$Mpc) is not completely negligible compared to the scale of the simulation itself ($500h^{-1}$Mpc). As a result, simulation cosmic variance could cause the empirical value of $\sigma_8$ to differ from its nominal value; in other words, the output amplitude of the density fluctuations (measured in cubical $31.25h^{-1}$-Mpc cells) might differ somewhat from the nominal value provided as input to the simulation. 

The nominal (input) value of $\sigma_8$ is 0.9 in the Millennium Simulation, and given this nominal value the linear bias is $b=1.15$ (see \citealp{Ising2} for bias determination). To obtain instead a directly measured (output) value of $\sigma_8$, we first determine the dark matter variance (in the cubical cells) of the simulation; we can then use linear theory (as implemented in CAMB\footnote{Code for Anisotropies in the Microwave Background \citep{CAMB}: \texttt{http://camb.info/}}) to determine the expected dark matter variance -- in the same size cubical cells -- given the nominal $\sigma_8$.\footnote{Note that the use of cubical rather than spherical cells affects the result by a few per cent; see \citet{ReppApdf} for the handling of this effect, as well as of power spectrum aliasing, etc.} The ratio of the two variances (actual vs. expected) will also be the ratio of the empirical and nominal values of $\sigma_8^2$, yielding a direct measurement of $\sigma_8 = 0.875$. Likewise, the directly measured value of $b^2$ is simply the ratio of the simulation galaxy and dark matter variances, yielding $b = 1.17$.

\begin{figure}
    \leavevmode\epsfxsize=9cm\epsfbox{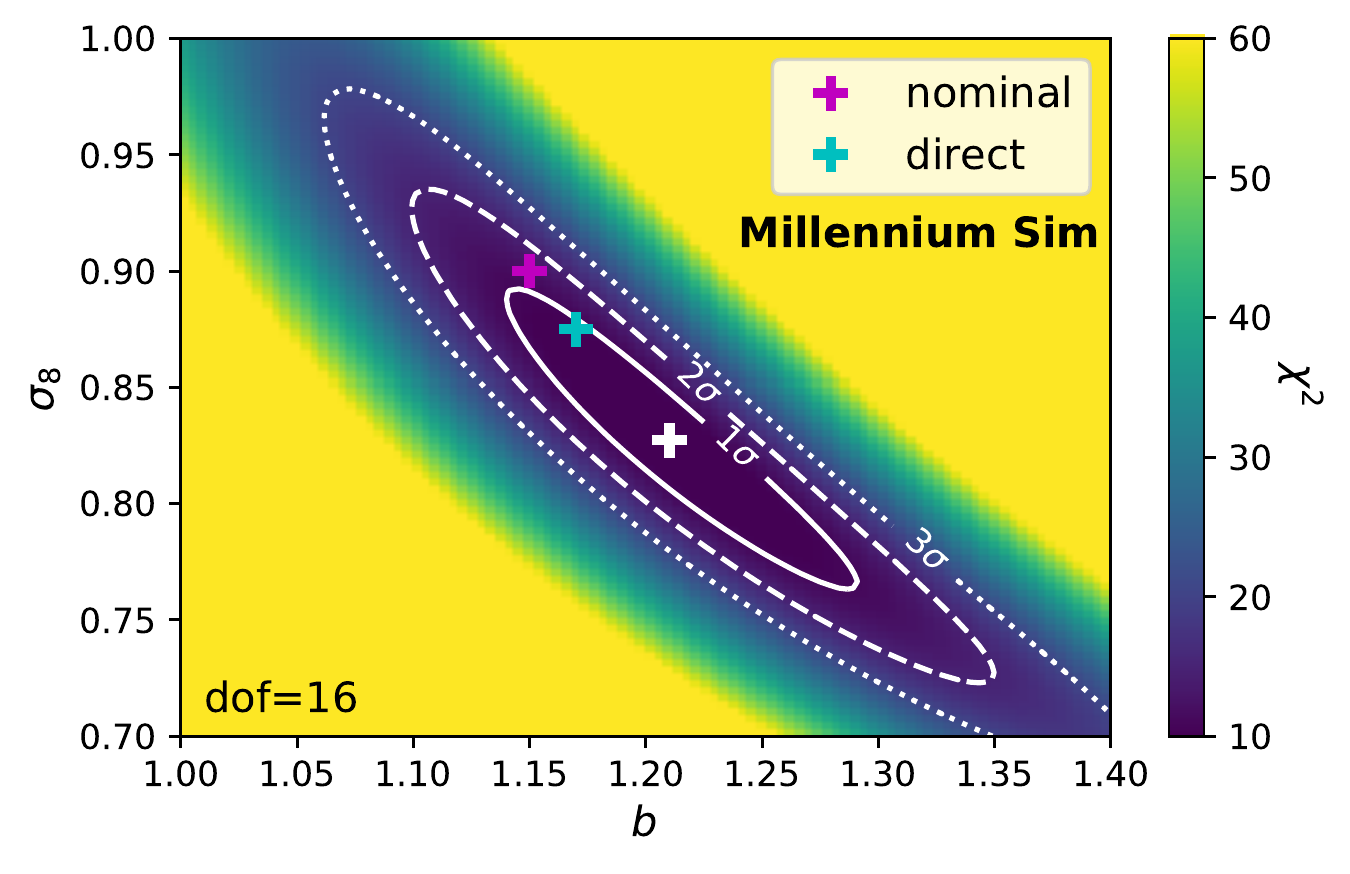}
    \caption{Values of $\chi^2$ from fitting $b$ and $\sigma_8$ to the redshift-space counts-in-cells probability distribution of Millennium Simulation galaxies at $31.25h^{-1}$-Mpc scales. The white, magenta, and cyan crosses show, respectively, the best-fit, nominal (input), and directly-measured (output) values for the simulation.}
\label{fig:MSfit}
\end{figure}

At this point we can apply the method of Sections~\ref{sec:pdffittheo} and \ref{sec:pdffitdesc} to simultaneously fit $\sigma_8$ and $b$ to the redshift-space Millennium Simulation PDF. We compute the $\chi^2$ for each combination of the two parameters to obtain the best-fit values. Confidence regions for $b$ and $\sigma_8$ appear in Figure~\ref{fig:MSfit}, which also shows  the nominal (simulation input) values in magenta and the directly-measured (simulation output) values in cyan.
The fit recovers the directly-measured galaxy bias and matter fluctuation amplitude to within $1\sigma$, with the dominant source of uncertainty being the error on the measured $\mathcal{P}(N)$.

We thus conclude that this method is capable (within the uncertainties inherent to the simulation PDF) of simultaneously recovering both $\sigma_8$ and the linear galaxy bias.

\section{SDSS Data}
\label{sec:data}
We now apply this technique to the the Main Galaxy Sample (MGS) from data release seven (DR7; \citealp{Abazajian2009}) of the Sloan Digital Sky Survey (SDSS; \citealp{York2000}). We obtain the galaxy catalog from the NYU-VAGC\footnote{\texttt{http://sdss.physics.nyu.edu/vagc/lss.html}} (New York University-Value Added Galaxy Catalog; \citealp{Blanton2005}). Restricting ourselves to the North Galactic Cap contiguous area, we follow \citet{Ross2015} in using the `safe0' catalog and in applying the following magnitude and color cuts: $M_r < -21.2$, $g-r > 0.8$. Furthermore, we restrict ourselves to the redshift range $0.07 < z < 0.17$, a somewhat smaller range than that of \citet{Ross2015}, in order to avoid the drop in number density apparent in their fig. 2. The result of these cuts is a roughly homogeneous, volume-limited sample of galaxies. For this analysis we assume the Planck cosmology \citep{Planck2018} except, of course, for $\sigma_8$.

To mimic our simulation analysis, we split the MGS into nine radially-concentric layers (each $31.25h^{-1}$Mpc thick), thus covering redshift-space $207.6h^{-1}$Mpc $\le s \le 488.9h^{-1}$Mpc. We tile each layer with approximately-square cells\footnote{Specifically, the declination intervals $\Delta \delta = \sqrt{\Omega}$, where $\Omega$ is the solid angle subtended by the cell.} bounded by lines of constant declination and right ascension, choosing their size to obtain cell volumes of $(31.25h^{-1}\textrm{Mpc})^3$.

\begin{figure}
    \leavevmode\epsfxsize=8.5cm\epsfbox{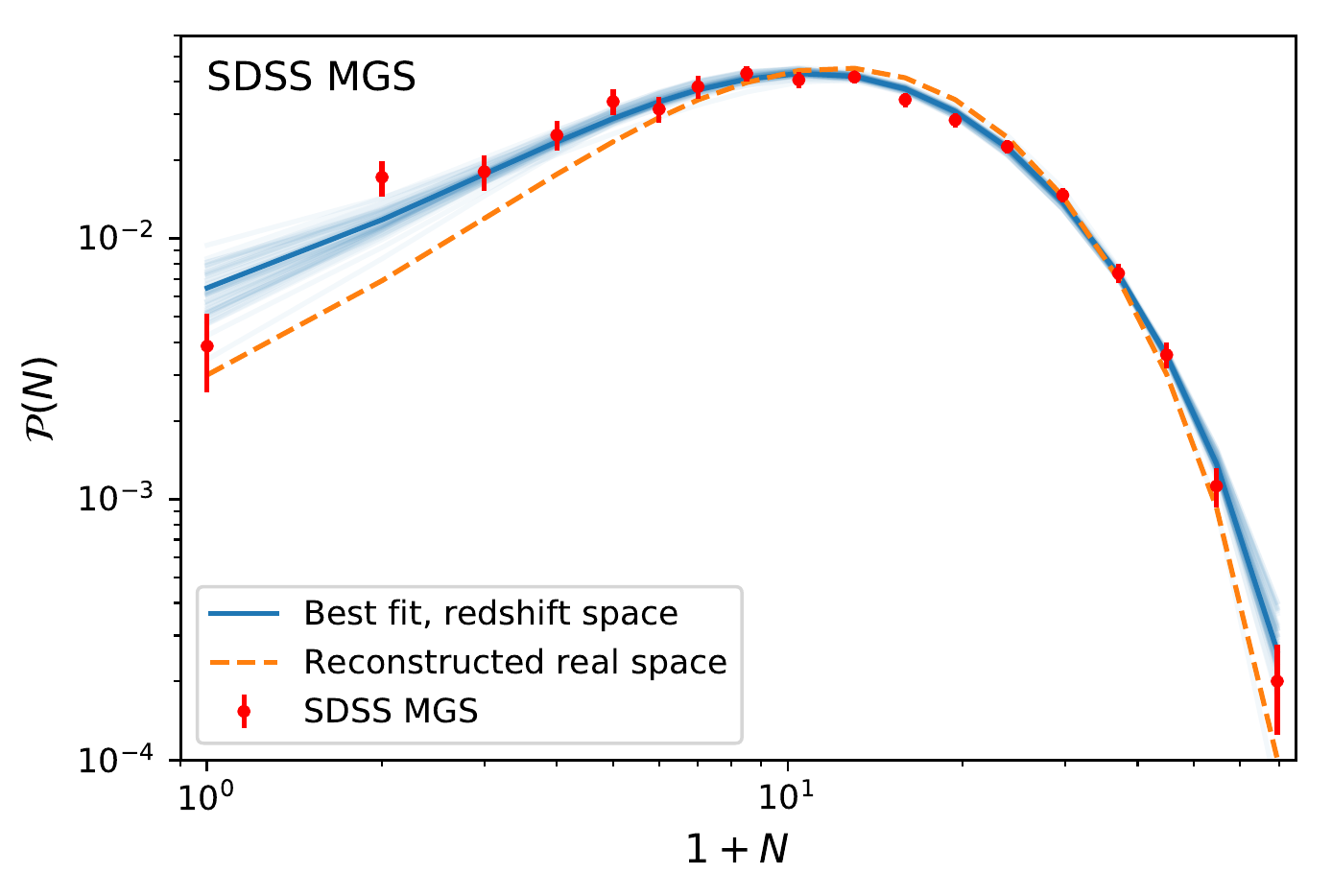}
    \caption{Measured and fit counts-in-cells probability distributions from the Sloan Main Galaxy Survey. Data points show the measured (redshift space) counts in cells; curves show both the best-fit redshift-space solution as well as the same solution pulled back into real space (using Equation~\ref{eq:sig8eff}). Light blue curves show fifty models randomly selected from the $(b, \sigma_8)$ parameter space according to likelihood.}
    \label{fig:PNs}
\end{figure}

We now must deal with the varying survey completeness of the different cells. To do so, we make use of the 20 random catalogs available from the NYU-VAGC; combining the catalogs, we obtain over 41 million random positions covering the region of sky which falls within the survey geometry but outside the bright star masks. We first discard any survey cell which is not at least 90 per cent complete. The remaining cells, though mostly complete, contain less effective volume than  $(31.25h^{-1}\textrm{Mpc})^3$ due to the masked-out areas. For these cells we expand the boundaries by the requisite amount in order to obtain the correct effective volume, since the CIC distribution is not sensitive to the shape of the cell \citep{Szapudi1998a}. Of course, the expansion might itself run into masked areas; furthermore, despite the large number of random points in the catalogs, there still exist Poisson fluctuations (albeit typically sub-per cent) in the number of points within each survey cell. We thus use a stochastic criterion for accepting the size of an expanded cell -- our acceptance rate is proportional to the probability that any deficit in random points is due to Poisson fluctuations, and otherwise we iterate the boundary expansion for that cell.

After this adjustment for completeness, we obtain 2,328 cells of effective volume $(31.25h^{-1}\textrm{Mpc})^3$ each, which together contain a total of 40,276 galaxies with median redshift $z_\mathrm{md} = 0.1397$. From these we obtain,  as in Section~\ref{sec:pdffitval}, the counts-in-cells PDF displayed in Figure~\ref{fig:PNs}.

\section{Results}
\label{sec:results}
\begin{figure}
    \leavevmode
    \epsfxsize=9cm
    \epsfbox{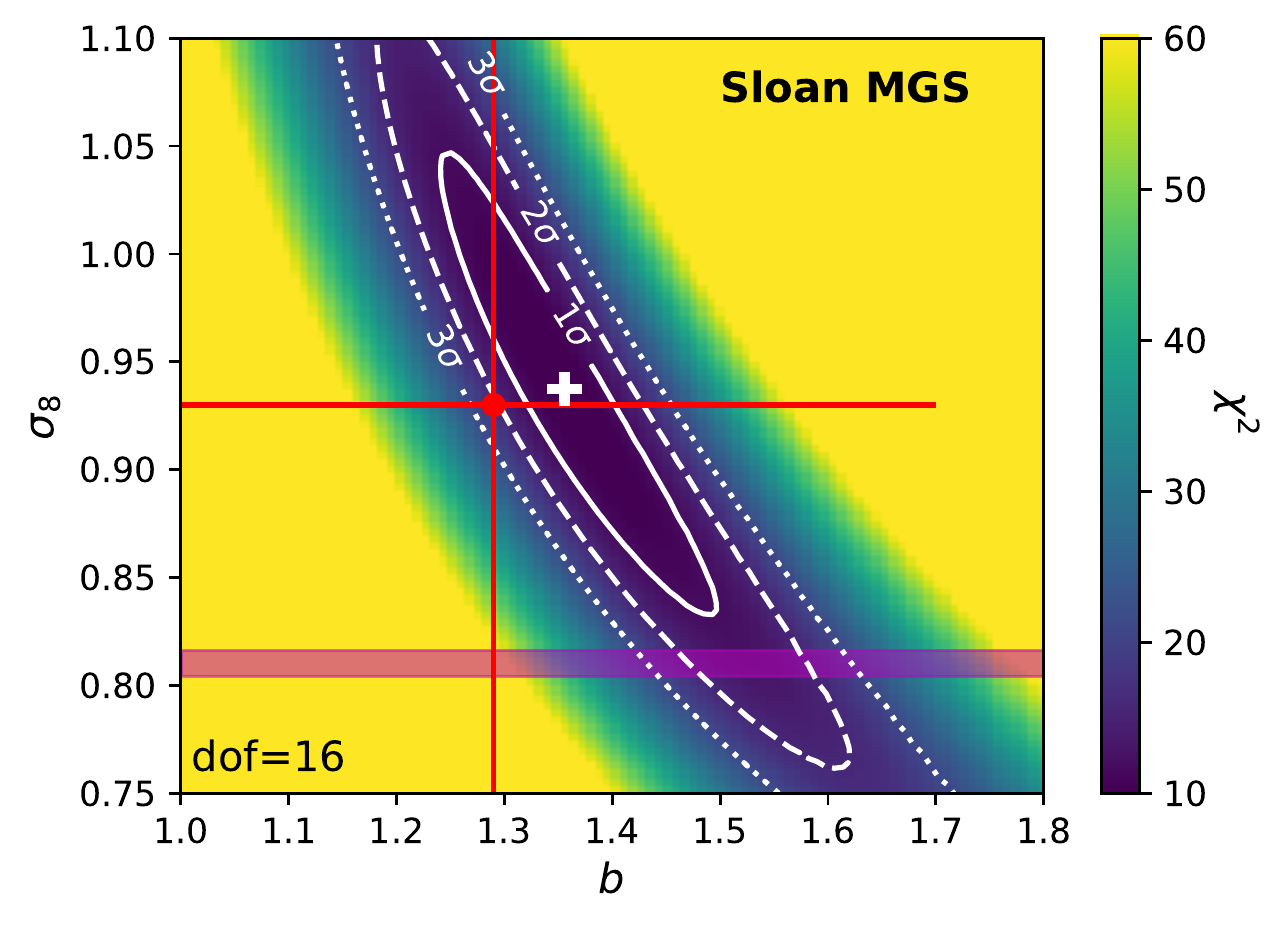}
    \caption{Values of $\chi^2$ from fitting $\sigma_8$ and galaxy bias $b$ to the SDSS Main Galaxy Sample counts-in-cells probability distribution. The white cross denotes the best-fit values, and the magenta band denotes the Planck (\citeyear{Planck2018}) $1\sigma$-confidence interval for $\sigma_8$. The results of \citet{Howlett2015} appear in red, having been determined from essentially the same galaxy sample using a different procedure.}
\label{fig:SDSS_fit}
\end{figure}

Using the model of Section~\ref{sec:pdffitdesc} we perform a joint fit of $\sigma_8$ and galaxy bias $b$ to the SDSS galaxies described above. Figure~\ref{fig:PNs} shows the best-fit probability distributions, and Figure~\ref{fig:SDSS_fit} shows the confidence regions for the parameters. In particular, at $z \sim 0.14$, we find $1\sigma$-confidence intervals of $\sigma_8 = 0.94^{+.11}_{-.10}$ and $b = 1.36^{+.14}_{-.11}$.

The magenta band in the figure shows the Planck (\citeyear{Planck2018}) $1\sigma$-confidence interval for $\sigma_8$; our result is consistent with it to about $1.5\sigma$.

The figure also shows the results of \citet{Howlett2015}, who measure redshift space distortion in the SDSS MGS and derive values of $f\sigma_8 = 0.49^{+.15}_{-.14}$ and $b\sigma_8 = 1.20 \pm .15$. If we take $f = \Omega_m^{0.55}$, the corresponding results from our fit are $f\sigma_8 = 0.49 \pm .05$ and $b\sigma_8 = 1.27 \pm .04$. Going the other direction, the \citet{Howlett2015} results translate into $\sigma_8 = 0.93^{+.29}_{-.27}$ and $b = 1.29^{+.43}_{-.41}$, which appear in red on Figure~\ref{fig:SDSS_fit}. Although we and \citeauthor{Howlett2015} use essentially the same data set, their method of analysis is complementary to ours, given that they extract information specifically from the redshift-space distortions instead of transforming them away, as we do. It is thus apparent that our result is consistent with their measurement as well.

We should note that \citet{Howlett2015} fit more parameters than we do, using a more sophisticated bias model as well as two additional parameters to capture variations in the background cosmology. The extra marginalization in this more complete fit is at least partially responsible for the higher uncertainties in their estimates. Since, as noted above, their analysis is complementary to ours, a combined analysis could potentially provide even tighter constraints. 
 
\section{Conclusions}
\label{sec:concl}
The variance of the one-point CIC probability distribution function captures most of the information on the power spectrum amplitude, while higher moments contain complementary cosmological information. In consequence, the galaxy CIC distribution in the mildly non-linear regime breaks the degeneracy between $\sigma_8$ and the linear galaxy bias $b$ that plagues power spectrum-only measurements.

Fitting parameters requires a prescription for the dark matter distribution; we use the GEV prescription of \citet{ReppApdf}. It also requires a means of unraveling the effects of redshift space distortion (Equation~\ref{eq:sig8eff}). Using a Millennium Simulation galaxy catalog in $31.25h^{-1}$-Mpc cubical cells, we verify that the PDF-fit technique successfully recovers the simulation-output values of $\sigma_8$ and $b$. Note that we perform our measurements on scales where linear theory for redshift distortion and bias is adequate, despite the fact that the CIC distribution function is mildly non-linear. On scales much larger, we would have less information in non-Gaussianity; on smaller scales, the linear approximation to bias and redshift distortions would break down.

We then apply the technique to the SDSS Main Galaxy Sample, again using $31.25h^{-1}$-Mpc survey cells, obtaining $\sigma_8 = 0.94^{+.11}_{-.10}$ and $b = 1.36^{+.14}_{-.11}$. These values are consistent with those of Planck (\citeyear{Planck2018}) as well as those derived (using a complementary method) by \citet{Howlett2015}. Note that this is the first time that the GEV prescription (Equations~\ref{eq:GEV}--\ref{eq:GEV_t}) for the distribution of dark matter has been tested with empirical data; these results indicate the reliability of this prescription on these scales.

We conclude that the theory of the galaxy CIC probability distribution function is mature enough that measuring it will be a significant source of information complementary to the power spectrum in the analysis of galaxy surveys.

\section*{Acknowledgements}
The Millennium Simulation data bases used in this work and the web application providing online access to them were constructed as part of the activities of the German Astrophysical Virtual Observatory (GAVO). This work was supported by NASA Headquarters under the NASA Earth and Space Science Fellowship program -- ``Grant 80NSSC18K1081'' -- and AR gratefully acknowledges the support. IS acknowledges support from National Science Foundation (NSF) award 1616974. 

\bibliographystyle{astron}
\bibliography{Thesis_Proposal}

\label{lastpage}

\end{document}